# Photocatalytic active ZnO$_{1-x}$S$_x$@CNTs heteronanostructures

Andjelika Bjelajac[a, b*], Ileana Florea[a, c], Mihai Zamfir[a, d], Sandrine Tusseau Nenez[e] and Costel Sorin Cojocaru[a]

[a]LPICM, CNRS, Ecole polytechnique, IP Paris, 91228 Palaiseau Cedex, France.

[b]Luxembourg Institute of Science and Technology (LIST), Maison des Matériaux, 28, avenue des Hauts-Fourneaux, L-4365 Esch-sur-Alzette, Luxembourg.

[c]CRHEA, CNRS, UMR7073, Rue Bernard Grégory, 06905 Sophia-Antipolis Cedex, France.

[d]National Institute for Laser, Plasma & Radiation Physics (INFLPR), Atomistilor Street, No. 409, Magurele, Ilfov RO-077125, Romania

[e]LPMC, CNRS, Ecole polytechnique, IP Paris, 91228 Palaiseau Cedex, France

*Corresponding author, andelika.bjelajac@list.lu

**Abstract**

Herein, we report on the use of vertically aligned multiwall carbon nanotubes (CNTs) films as support for ZnO/ZnS photocatalytic active nanostructures. The CNTs were synthetized via a hot-filament chemical vapor deposition (HfCVD), using Fe catalyst on top of Al$_2$O$_3$ buffer layer. Controlled point defects in the CNTs outer walls were created by exposure to a low pressure nonthermal water vapors diffusive plasma and acted as seeds for subsequent pulsed-electrodeposition of Zn nanoparticles. This was to achieve a direct and improved contact between the nanoparticles and CNTs. To obtain ZnO, ZnS and mix phase of ZnO/ZnS spread on CNTs, the oxidation, sulfurization and 2 steps subsequent annealing in oxygen and sulfur rich atmospheres were applied. High resolution transmission electron microscopy (HRTEM) with energy dispersive X-rays spectroscopy (EDS) in scanning mode, provided the chemical mapping of the structures. X-ray diffraction (XRD) analyses proved the hexagonal phase of ZnO and ZnS, obtained after oxidation in H$_2$O and S vapors, respectively. In the case of the samples obtained by the 2 steps subsequent annealing, XRD showed mainly the presence of ZnO and a small amount of ZnS. The benefit of the secondary annealing in S vapor was seen as absorption enhancement of the ZnO$_{1-x}$S$_x$@CNTs sample having the absorption edge at 417 nm, whereas the absorption edge of ZnO@CNTs was 408 nm and of ZnS@CNTs 360 nm. For all the samples, compared to the bare ZnO and ZnS, the absorption red shift was observed which is attributed to the CNTs involvement. Therefore, this study showed the double sides benefit to induce the absorption of ZnO of the visible light, one from S doping and second of CNTs involvement. The absorption enhancement had a positive impact on photocatalytic degradation of methyl blue dye, showing that ZnO$_{1-x}$S$_x$@CNTs heteronanostructure was the best photocatalyst among the studied samples.

Keywords: photocatalysis; nanostructures; electron microscopy; carbon nanotubes;

## 1    Introduction

Water waste treatment has been and remains an important issue that humanity is facing. One of the well-known strategy for water purification is photocatalysis [1]. When a photocatalyst absorbs the solar radiation, free carrier charges are generated, electron going from the valence to the conduction band thus leaving behind a positive "hole" which can be used for further subsequent reactions. Typical photocatalytic materials are inorganic semiconductors among which ZnO [2], [3] and ZnS [4–6] are intensely explored. Both of them are bio-safe and biocompatible materials and can be directly used in heterogeneous photocatalysis. A drawback relates to their optical absorbance being limited to the UV range, $E_g$ (ZnO) = 3.2 eV (387 nm) [7] and $E_g$ (ZnS) = 3.73 eV (332 nm) [8]. Furthermore, ZnO is unstable in water, i.e hydrolyzes over time to form Zn(OH)$_2$ on ZnO exposed surface thus rendering it catalytically inactive over time [9], [10]. Moreover, ZnS photodecomposes and undergoes self-oxidation in water with formation of excited holes, ZnS→ Zn$^{2+}$ + S [11]. One possibility to alleviate the photocorrosion is to use a hole scavenger such as methanol, KBr or Na$_2$S. However, as these additives are consumable, they would need to be supplied constantly. One approach to improve the photocatalytic performance of ZnO and ZnS consist in their coupling with carbon nanotubes (CNTs) to form a hybrid nanoheterostructured materials [12], [13]. CNTs have exceptionally high charge mobility and they are considered the ideal electron acceptor. This can trigger enhanced interfacial electron transfer from the photocatalyst to the CNTs and restrain the electron/hole ($e^-/h^+$) pair recombination [14], [15]. Additionally, grafting the photocatalyst on CNTs surface will prevent the agglomeration of the catalyst in the solution which negatively affects the photocatalytic performances [16] whilst at the same time the available photo-catalytic specific surface area is increased. Furthermore, the photocorrosion process is prevented by inhibiting the accumulation of the charge carriers within the catalyst [8]. On the other hand, the photocatalytic properties of ZnO can be improved by its doping with S [17] and/or by forming ZnS@ZnO core-shell heterostructure, which will extent the absorption in visible part of the spectrum and improve catalysts stability [17–21]. The strategy we propose is to decorate CNTs first with ZnO particles and then introduce S either as dopant into the ZnO structure or as a new outerlayer of ZnS onto ZnO. This approach will take advantage of the fact that the conduction band of ZnS lies at a more negative potential than that of ZnO, while the valence band of ZnO is more positive than that of ZnS [22]. Therefore, under visible-light irradiation, one can expect that the photogenerated electrons will transfer from the conduction band of ZnS to the conduction band of ZnO and together with photo-excited electrons from ZnO will be further accepted by the CNTs. Simultaneously, hole transfer is expected to occur from the valence band of the ZnO to the valence band of the ZnS. Subsequently, one can thus expect enhanced charge carrier separation and an improved photocatalytic activity. The doping of ZnO with S has already been proposed by Zhang et al. [23] where they synthetized ZnO nanoflowers via hydrothermal methods and a simple thermal treatment at 60 °C in 0.5 M $Na_2S$, resulting in a visible range responsive catalyst with 550 nm absorption edge. However, in their report, the pure S:ZnO exhibited poor photodegradation activity, which was explained as a consequence of the easy recombination of the photogenerated charges in S–ZnO. By additional Ag loading, which captured photoexcited electrons from S-ZnO, the photoactivity of the catalyst was shown to be improved. Herein, we propose the use of CNTs for the same purpose as Zhang et al. used Ag. This study aims thus to explore the contribution of CNTs on optical properties of the deposited Zn further oxidized and/or sulfurized to obtain ZnO and/or ZnS. Additionally, the effect of S incorporation in ZnO, by low temperature annealing in S vapors, was investigated. Therefore, the objective of our work was to demonstrate the two-side beneficial role on the photocatalytic activity of the ZnO@ZnS/CNTs heterostructure: i) involvement of CNTs to reduce the photogenerated charges recombination and to provide high surface area as a support for the catalyst, ii) doping of ZnO with S and creating ZnO@ZnS structure to increase the stability of the catalyst as well as absorption range.

## 2 Methods

### 2.1 Preparation of the catalyst for CNTs growth

High-purity aluminium or copper foils (Alfa Aesar 99.99%) were cut into pieces 5 cm×6 cm in size to be used as substrates for CNTs growth. The foils were first cleaned in acetone followed by isopropanol in an ultrasonic bath for 15 min. Subsequently, 30 nm $Al_2O_3$ buffer layer followed by a 5 nm Fe layer were deposited in a home-made UHV e-beam heated, molecular beam evaporation system. Prior to the deposition, the substrate was heated at 300 °C to ensure better adhesion of the deposit to the substrate and the same temperature was maintained during the double layer deposition process.

### 2.2 CNTs growth

The Fe catalyst precursor reductive pre-treatment and the CNT subsequent growth were conducted in a homemade double hot filament chemical vapor deposition (d-HFCVD) reactor. The set-up and process details have been reported previously [24] and thus, herein we provide just a brief description of the procedure. The reductive pre-treatment under activated hydrogen was done to induce the formation of small non-oxidized Fe metal nanoparticles from the Fe thin film. These Fe nanoparticles served as catalysts for CNTs growth. The pre-treatment step is immediately followed by the CNT synthesis, by introducing methane ($CH_4$) at 50 standard cubic centimeters per minute (sccm) flow rate, whereas the hydrogen flow rate was modified from 75 to 20 sccm. The power of the two filaments for $CH_4$ was set to be 450 W, whereas for the $H_2$ filaments was 550 W. The overall pressure was maintained to 12 mbar. These growth conditions are maintained during 30 minutes, after which the chamber was evacuated and the samples removed from the hot zone. [25]

A dense black coatings on the upper side of the substrates with Fe/Al₂O₃ catalyst were observed, which is an indication of a successful synthesis of vertically aligned CNTs carpet (VACNTs) [24].

After the growth of the CNTs, using a home made DC plasma enhanced chemical vapour deposition reactor (PECVD) system [26] the samples were exposed to a low pressure nonthermal water vapors diffusive plasma in order to induce excessive amount of point defects in the CNTs outer walls structure. These defects can be further used as preferential nucleation sites for Zn during the electrodeposition step. For the nonthermal plasma treatment the water vapor gas was injected in the PECVD chamber at a flow rate of 10 sccm and a pressure of 1.6 mbar. The temperature of the sample electrode gradually increased in time upon plasma exposure, i.e. for 3 minutes of plasma treatment the temperature reached 70 °C. After the ignition of the plasma, the applied voltage to the anode was maintained quasi constant in a range of 30-45 V with a current intensity between the two multi holes electrodes of 0.22-0.24 A and without applying power on the sample electrode.

## 2.3    Zn deposition, oxidation and sulfurization

Zinc nanoparticles were deposited on CNTs sidewalls via pulsed-electro-deposition (PED) process using a three electrodes setup, with a Ag/AgCl electrode as a reference electrode, the CNTs sample as a working electrode and a graphite counter electrode. The setup was power supplied via a Bio-Logic potentiostat and EC-lab software for setting the electrodeposition parameters and designing the pulses shape applied on the working electrode. The electrodeposition process was performed in a Watts bath solution consisting of a mixture of 143.78 gl⁻¹ ZnSO₄·7H₂O, 30 gl⁻¹ H₃BO₃ and 0.5 gl⁻¹ ascorbic acid (C₆H₈O₆). The acids were added in order to buffer the electrolyte acidity as to not damage aluminium and aluminium oxide layer. For the electrodeposition of Zn nanoparticles, pulses of 40 ms with a potential amplitude of -7.5 V followed by a 120 ms resting time in an open voltage circuit were applied. The total number of pulses was set at 500 for all the fabricated samples. A more detailed description of the electrodeposition technique is reported elsewhere [27].

Subsequent to the Zn NPs deposition on CNTs, part of the obtained samples was annealed in H₂O vapor, to obtain ZnO, whereas another part was annealed in S vapor to obtain ZnS. The heating regime was in both cases the same with 10 °C/min rise to 500 °C and anneal for 30 min at 3 mbar. To introduce S in ZnO/CNTs structure, a secondary sulfurization step was applied at a lower process temperature of 250 °C, 10 °C/min, for 30 min at 3 mbar (third type of the studied heterostructure).

For more comprehensive explanation of the experimental path for the samples' preparation, a corresponding schematic is provided in Fig. 1.

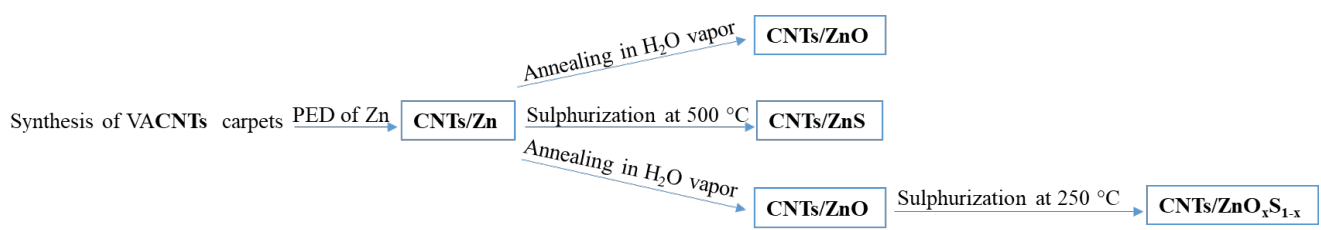

**Figure 1**. Schematic of experimental path done in this work

## 2.4    Characterization Techniques

Scanning electron microscopy (SEM) with energy dispersive spectroscopy (EDS) analyses were carried out on a HITACHI S 4800 microscope at 10 kV.

Morphological and structural analyses were performed using an ThermoFisher TITAN Themis transmission electron microscopy (TEM) operating at 300 kV. For all observation, prior to the analysis, the surface of the samples were scratched with a diamond scriber and fragments were collected on the carbon grid. No solvent was used for the preparation in order to preserve the as-synthetized morphology of each analyzed sample. The chemical analyses were performed via EDX in scanning transmission mode (STEM) of the electron microscope, using an X-MAX Oxford detector. The experimental conditions were set so that the total current within the probe used for the STEM-HAADF EDS chemical analysis was about 85 pA. For the STEM-HAADF-EDS chemical mapping we selected characteristics x-ray energies for carbon with Kα = 0.277 keV, oxygen with Kα = 0.523 keV, zinc with Kα = 8.639 keV and sulfur with Kα = 2.308 keV.

The STEM-HAADF images were recorded using a camera length of 110 mm where the intensity of the recorded STEM-HAADF image reflects variation of the average atomic number Z of the probed element.

Grazing incidence X-Ray diffraction (GIXRD) measurements were carried out using a high-resolution diffractometer (Smartlab, Rigaku) equipped with a high-flux 9 kW rotating anode (copper source), the CBO unit and the HyPix-3000 high-energy-resolution semiconductor detector. No monochromator was used in order to maximize the collected intensity. The optical configuration was as follows: parallel beam, incident slit IS = 1mm, 2.5° Soller slits to limit the axial divergence beam and parallel slit analyzer (PSA = 0.228°) to limit the equatorial divergence beam. The incidence angle ω was set from 1.5° to 4° so that the beam penetrates the thickness of the deposited film. The angular range for data collection was from $2\theta$ = 12 to 80 $2\theta_{Cu}$ °, with a step of 0.03° and a scan speed of 5°/min or 0.5°/min for accurate data. GIXRD patterns are plotted in log scale.

The crystallite size was estimated using the Halder-Wagner method [28]. The intercept of the plot of $(\beta/d^*)^2$ versus $\beta/(d^*)^2$ gives the mean strain value $\eta$ and the slope the mean apparent size, $\varepsilon$ :

$$\left(\frac{\beta}{d^*}\right)^2 = \frac{1}{\varepsilon} * \frac{\beta}{d^{*2}} + \left(\frac{\eta}{2}\right)^2$$

With $d^* = 1/d = 2\sin\theta/\lambda$, $\beta$ the broadening of the peak and $\eta = \langle\varepsilon_L^2\rangle^{1/2}$ i.e. the root-mean-square strain value.

The volumetric apparent size $D_v$, in the spherical grain approximation, is calculated as $D_v = 4/3\eta$ [29]. With our optical configuration, the instrumental broadening is fixed by the aperture of the analyzer, its value is constant over the entire angular range. From the pattern of $Al_2O_3$ standard powder the full width at half maximum (FWHM) was taken to be 0.220° and the integral breath (IB) 0.262°. The observed Lorentzian width is the sum of the width due to particle size of the sample and instrumental broadening, while the square of the observed Gaussian breadth is given by the sum of the squares of the instrumental and the sample strain Gaussian widths.

When not enough peaks were available to plot a linear regression, the crystallite was only estimated using the Scherrer formula, where the broadening of the diffraction peak was considered due to the size effect and the constant K was taken equal to 1.07 (considering IB for the broadening, as done for the Halder-Wagner method).

Diffuse reflectance spectra (DRS) were recorder on a Peker Elmer Lambda 950 nm UV-Vis-NIR spectrophotometer.

## 2.5 Photocatalytic measurements

The photocatalytic activity of the samples was tested by monitoring the photodegradation of methyl blue dye in time with and without the catalysts. A 30 cm distance between the catalyst and the Xe lamp, operating at 50 W, was set. The 50 mL of dye (starting concentration $c_0$ = 20 mg/L) was irradiated for 6 h to measure the baseline. The catalysts were then sequentially immersed in a fresh solution of the same initial concentration and a first sampling was performed after 30 mins in dark to investigate the change of the concentration due to simple adsorption. The influence of irradiation over time was monitored by sampling of the irradiated dye with a catalyst immersed after 30, 60, 120, 240 and 360 min, respectively. The Varian Cary® 50 UV-Vis Spectrophotometer was used to measure the absorption of the dye solutions which is taken as a change in dye concentration. As a measure of degradation, the decrease in concentration is determined by monitoring the peak at 662 nm. To be able to compare the effectiveness of the catalyst, all films were cut to have the 2 $cm^2$ geo. surface.

## 3 Results and Discussion

### 3.1 Morphological and chemical analyses

Fig. 2 a) shows typical SEM images where the morphology of as-synthetized 45 μm long and vertically aligned CNTs is illustrated. From the TEM image illustrated in inset of Fig. 1a) one can see that these multiwall CNTs have 4-5 walls with 5-6 nm sized inner diameter, estimated by a manual measurement of 235 MWCNTs. The corresponding histogram is provided in the Supplementary file, as Fig. S1. This morphology is expected for the growth by using 5 nm Fe nanoparticles catalyst in our chemical vapor deposition (CVD) set-up. The subsequent deposition of Zn was uniform all along the CNTs outer walls which is clearly visible on Fig. 2 b). The particles were in 50 to 280 nm range and most of them were in the shape of pellets with a hexagonal base. To determine their chemical composition and their precise phase, the XRD analyses were performed. The peak at 72.5° $2\theta$ detected by XRD confirms that the deposit was pure Zn with a very small amount of ZnO hexagonal phase (Fig. 2 c)). The observed oxidation could be a consequence of oxygen exposure both from the electrolyte and from the air. Using the Halder-Wagner analysis, the volumetric apparent crystallite size of Zn is 20 nm, whereas the Scherrer formula gives an average size of 23 nm.

Two very broad peaks indexed at 33.1 and 59.5° $2\theta$ can be explained by the presence of iron oxide(s). This oxidation of iron means that not all the catalytic iron served for the CNTs growth, but still some of it remained on the substrate.

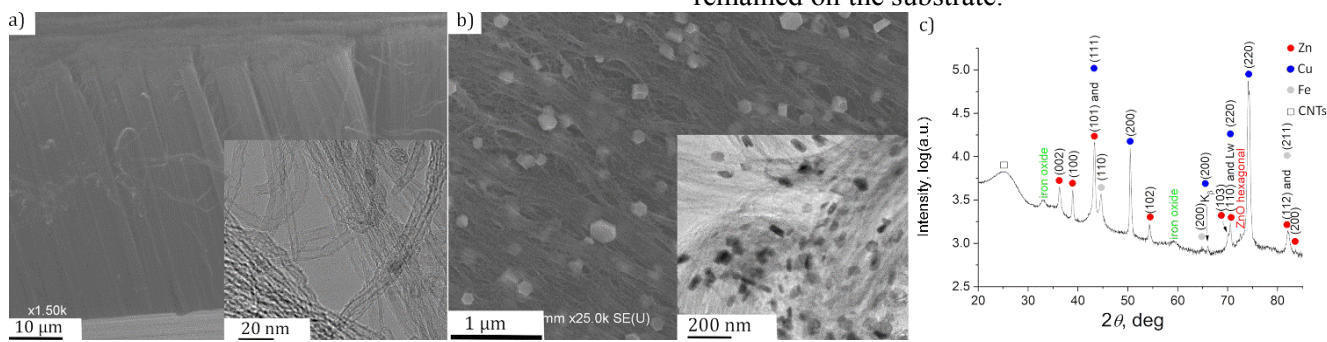

**Figure 2.** a) SEM image of as-synthetized CNTs and the TEM micrograph in the inset, b) SEM image of CNTs after PED of Zn and the corresponding TEM micrograph in the inset, and c) GIXRD pattern of CNTs after PED of Zn

Using high resolution (HR-TEM) imaging (Fig. 3), the presence of smaller particles deposited all along the CNTs outer walls was observed with an estimated size between 5 to 12 nm. A detailed analysis of the nanoparticles allowed the determination of the interplanar distances $d_1 = 0.28$ nm (100), $d_2 = 0.25$ nm (101) and $d_3 = 0.26$ nm (002) which are specific to the ZnO hexagonal phase. The analyzed particle presented in Fig. 3b) is covered by a ~10 nm thick layer. Furthermore, EDS analysis was performed in STEM mode and the results are resumed in Fig. S2 in the supplementary file together with the chemical composition of the analyzed fragment given in Table

S1. What is observed is that oxygen is mostly present on the surface of CNTs, where there was no signal identification for zinc component. To further investigate this, the point EDS analysis was performed in different areas within the samples which revealed us that the amount of oxygen was always higher than the zinc (see also in supplementary file Fig. S3). We assume that since the plasma treatment, applied after CNTs synthesis, induced defects on the surface of CNTs and those defects can also trap oxygen molecules ($O_2$) as well as hydroxyl or carboxyl groups.

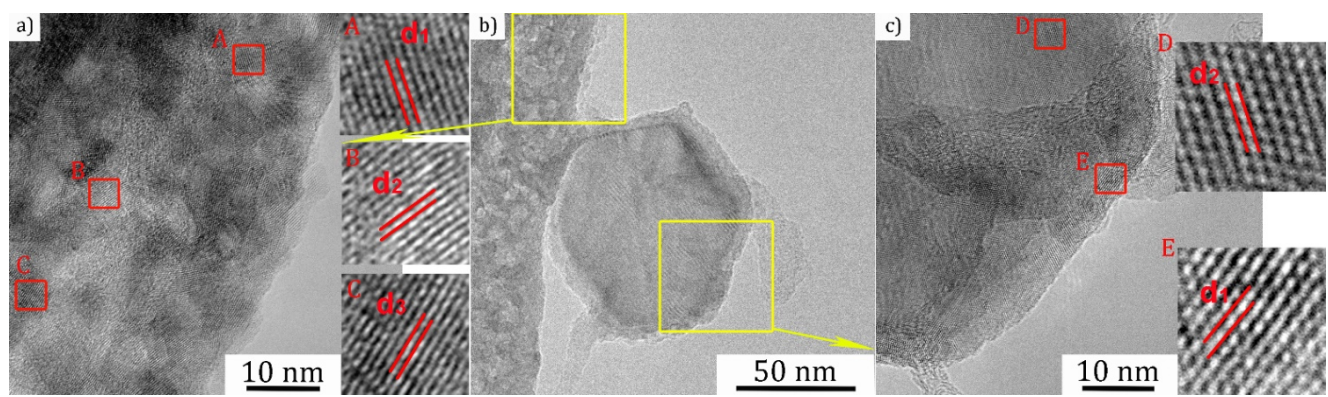

**Figure 3.** HRTEM micrographs of Zn@ZnO deposited on CNTs: a) the fragment of the deposit consisted of agglomerated nanoparticles of 5-12 nm range, b) a hexagonal shaped ~50 nm particle with a zoom given in c) to present the HR of measured interplanar distances, d

To ensure the complete oxidation of Zn@ZnO on CNTs, the annealing in $H_2O$ vapor was performed (Fig. 1) and the characterization results are summarized in Fig. 4. As it can be observed from the SEM image (Fig. 4 a) the overall morphology of the deposit remains the same. All the particle preserved their hexagonal pellet-like morphology after the complete oxidation and present a bimodal size distribution. From the XRD analysis (Fig. 4 b) we observed that the obtained ZnO maintains its hexagonal phase structure (zincite, P63mc. JCPDS 00-036-1451). Using the Halder-Wagner analysis, considering that the instrumental broadening corresponds to the broadening of the Al peak (integral breath, IB = 0.2298 °), the volumetric apparent crystallite size of ZnO is 9.6 nm accounting for a small amount of microstrain (0.16 %) (the Scherrer formula gives an average size of 15 nm). HRTEM analysis also confirmed ZnO hexagonal phase, having measured the interplanar distances to be 0.28 nm that corresponded to the (100) plane of hexagonal ZnO (A-D).

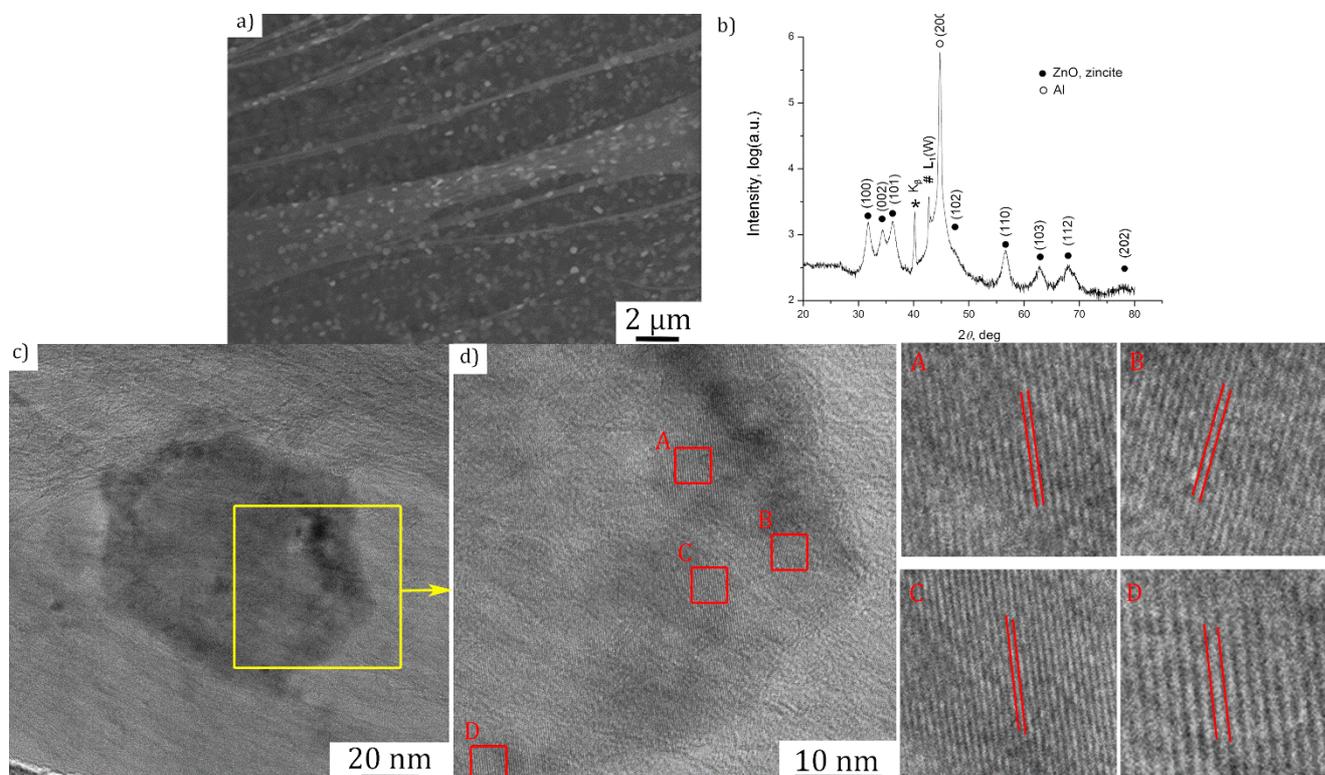

**Figure 4.** a) SEM, b) TEM micrographs, c) GIXRD pattern of ZnO on CNTs, obtained after annealing in $H_2O$ vapor. c) HRTEM micrograph of a hexagonal shaped particle and d) zoom of the area marked with yellow given to present the measured interplanar distances

For ZnS/CNTs heterostructure preparation, a high temperature sulfurization of as-deposited Zn@ZnO/CNTs structure was performed (Fig. 1). The morphological and chemical characterization results are summarized in Fig. 5. From the SEM images illustrated in Fig. 5 a) we discovered that due to sulfurization the particles have increased their size and changed their morphology. For this system we have distinguished again a bimodal distribution with nanoparticles presenting sizes between: i) 8-18 nm and ii) 180-330 nm. Regarding their morphology we observe that almost all particles have lost their initial hexagonal shape which turned to be more spherical after sulfurization. From the HR-TEM images we could measure the interplanar distances $d_1 = 0.31$ nm (100) and $d_2 = 0.33$ nm (002) specific to the ZnS component with a hexagonal phase. These results were confirmed by the XRD analysis presented in Fig. 5 b), where the hexagonal (wurtzite-6H, hexagonal, P63mc. JCPDS card 01-089-2739) phase of ZnS deposited on CNTs supported on Al foil (sharp peaks at 44.69°,

65.05 and 78.21° 2θ) can be identified. Using the Halder-Wagner analysis, the volumetric apparent crystallite size of ZnS is 7 nm with no microstrain. This is in agreement with TEM observation as it corresponds to the first bimodal size distribution. The second one probably corresponds to an agglomeration/aggregation of these crystallites into spherical nanoparticles.

The corresponding STEM-HAADF-EDS analysis of ZnS/CNTs (Fig. S4 in supplementary file) shows that the S content is a bit higher than that of Zn, which is in accordance with EDS results of as deposited Zn@ZnO, having higher amount of O than Zn (Table S2). This was expected since O ions were exchanged with S and the sulfurization parameters were optimized to get minimum amount of O thus, as much as possible pure ZnS on CNTs.

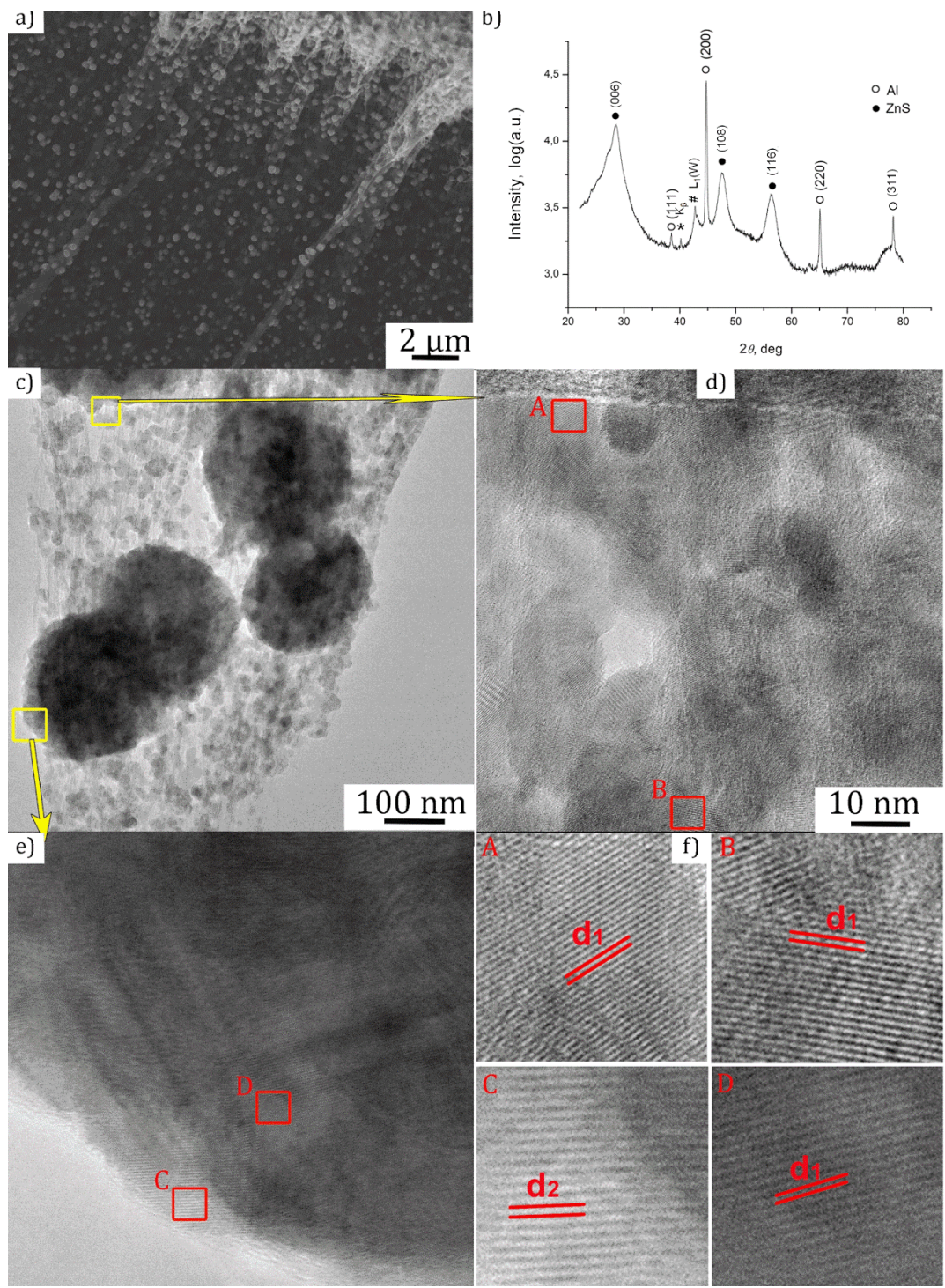

**Figure 5.** a) SEM micrograph, b) GIXRD pattern of ZnS on CNTs, obtained after sulfurization of ZnO as deposited on CNTs, c) TEM micrograph with 2 squared areas zoomed in d) and e) to present the measured interplanar distances

The ultimate goal of this research was to obtain the mixed phase ZnO@ZnS, and to investigate the form in which it resulted after the secondary annealing of ZnO/CNTs in S vapor (firstly annealed in $H_2O$ vapor, see Fig. 1). As seen from Fig. 6 the annealing in S vapor did not change the hexagonal pellet-like morphology of the particles. The recorded XRD pattern shows clearly the ZnO phase (hexagonal, P63mc. JCPDS card 00-036-1451), where the crystallite size is estimated to be 12 nm using Halder-Wagner plot. The small broad peak at $2\theta$ of 52.30° could be attributed to ZnS wurtzite phase (JCPDS card 01-089-2739), but all the other peaks were overlapping with very broad peaks of ZnO. The Scherrer formula gives a crystallite size about 65 nm. What is more, HRTEM measurement of interplanar distances performed on several fragments showed a mismatch with the crystallography data corresponding to ZnO. However, there were fragments with $d_1 = 0.310$ nm, which may be assigned to ZnS. Still, some fragments with $d_2 = 0.285$ nm were observed, which is attributed to (100) plane of ZnO zincite. Therefore, further STEM/EDS analysis had to be performed and the EDS analysis confirmed the presence of S in a ratio of at. % Zn:O:S ≈ 3:2:1 (see Supplementary file, Table S3). However, at closer look (Fig. 7), it appears that S is spread everywhere, not that localized to the analyzed particles whose signal for Zn and O are more intense than aside. Therefore, we can reasonably assume that the big particles are indeed ZnO with an outer cover layer of ZnS. Moreover, it is to be expected that the small particles present all along CNTs can more easily be sulfurized. Similar finding was reported by Ahn et al. [30] where they reported ZnS nanoparticles being spontaneously formed during the sulfidation of ZnO at a certain reaction temperature. The ZnS grains were primarily formed at the surface of ZnO because the sulfidation began from the top surface exposed to S source. Furthermore, Cheng et al. demonstrated the ZnO@ZnS heterostructure synthesis, also implying that ZnS is firstly being formed at the surface of ZnO, and then some part of S diffuses to the core of ZnO [31]. This is why smaller ZnO were more easy to sulfurize and those very small ZnS are difficult to detect on XRD and distinguish from bigger ZnO particles with probable ZnS outer layer.

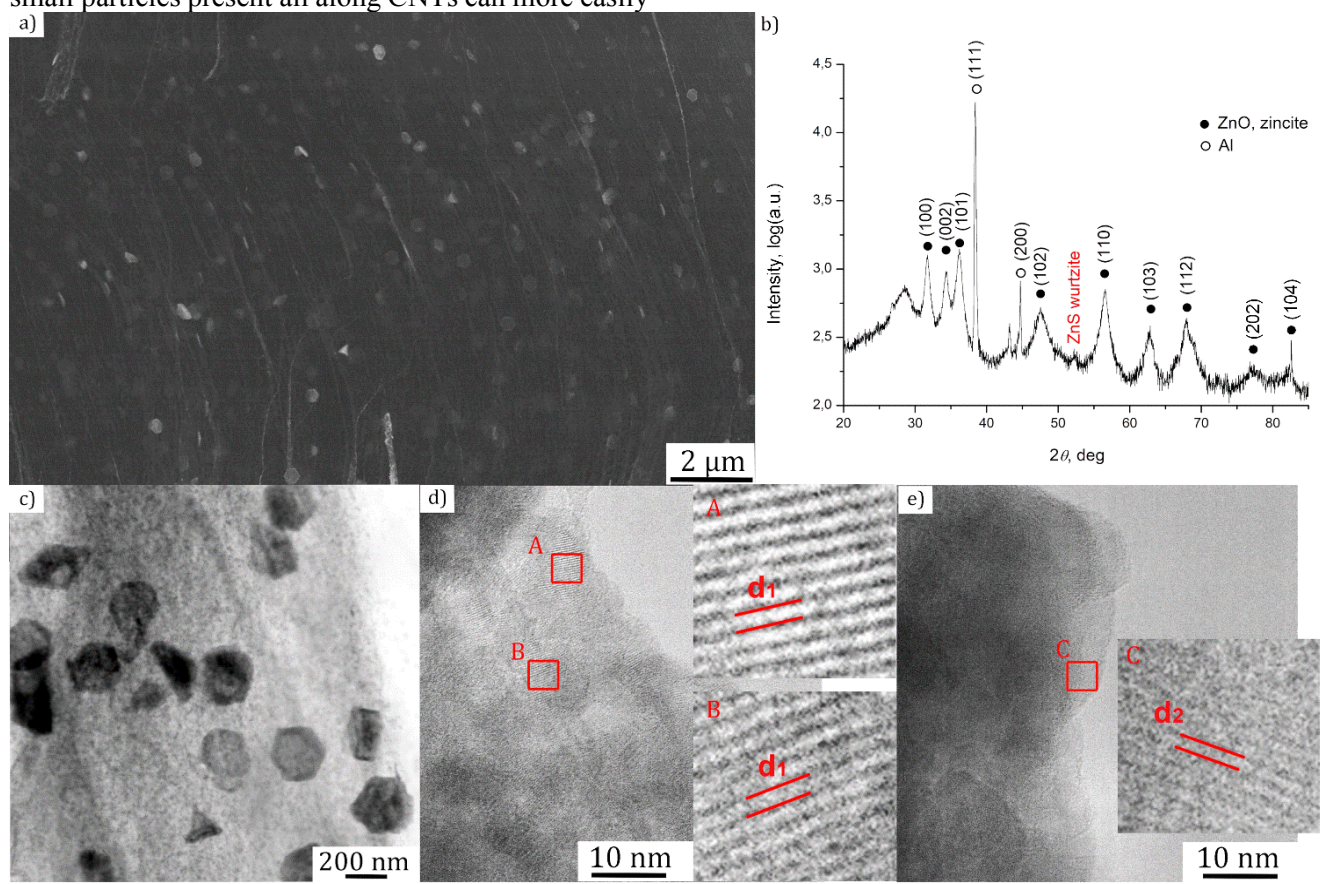

**Figure 6.** a) SEM micrograph, b) GIXRD pattern, c) STEM image of ZnO/CNTs after annealing in S vapor, d) and e) HRTEM micrographs with the measured interplanar distances

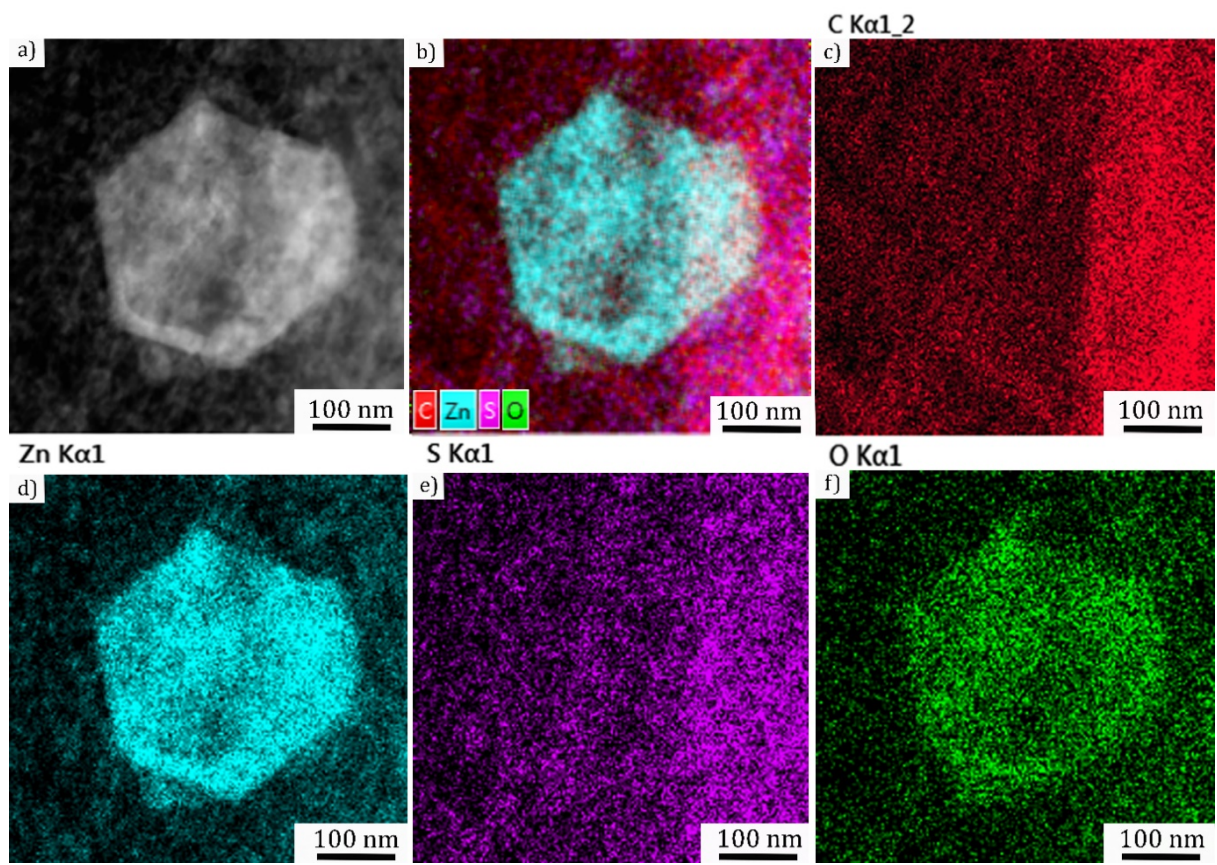

**Figure 7.** STEM-HAADF EDS chemical analysis of an area specific to the ZnO/CNTs structure after annealing in S vapor: a) STEM-HAADF image and b) its corresponding chemical mapping with the C component in red c), the Zn in blue d), the S in violet e), and O component in green f)

## 3.2    Photocatalytic activity

In order to explore the differences in optical properties of the synthesized samples, the DRS spectra were recorded. Using the transformed Kubelka Munk function $F(R) = (1-R)^2/(2R)$, where R is the measured reflectance in %, the Tauc plot is used to estimate the bang-gap energies of the heterostructures (Fig. 8) [32]. The CNTs film showed an absorbance in wide range, up to 765 nm. It is expected that multi-wall CNTs cannot produce the free photogenerated electrons and therefore they cannot be used as photocatalysts by themselves [33]. However, CNTs can have a positive effect on the optical properties of ZnO and ZnS, seen as the narrowing of the band gap, since for the bare ZnO, the band gap is 3.2 eV nm and for the bare ZnS 3.73 eV. For the obtained ZnO/CNTs and ZnS/CNTs films the estimated band gaps were: 2.98 eV and 3.35 eV, respectively. Similar finding was reported by Lonkar et al. where they showed the beneficial role of graphene on the absorption properties of ZnO-ZnS heterostructures [34]. Incorporated graphene served as macromolecular sensitizer that caused the narrowing of the bang gap of ZnO-ZnS heterostructures compared to the samples without graphene. Moreover, we can emphasize the marked additional enhancement of absorbance of the S-ZnO/CNTs sample. We can infer that the S incorporation in ZnO induces a narrowing of the band gap, by mixing the S p states with O p states [35]. Therefore, this study highlights the two sides benefit, one originating from CNTs as a support and another from the S doping of ZnO.

Other reported values for absorption edge of ZnS/CNTs heterostructures are either shifted to lower wavelength, which is attributed to quantum size of ZnS nanoparticles [28, 29],[38] or are remaining the same as for bulk ZnS [14]. The observed red shift of absorbance in our sample is to be related to the bimodal size distribution of ZnS particles, where bigger ones are in range 180-330 nm. This does not mean that we do not have small nanoparticles in possible quantum regime, but the contribution of CNTs incorporation is more evident, as previously reported for similar structures, i.e. where graphene served as photosensitizer of ZnO-ZnS heterostructures [34]. We assume that we have the synergistic effect of ZnS-ZnO/CNTs hybridization that promoted the band gap narrowing compared to the bare ZnO and ZnS. What is more, the ZnO/CNTs structure obtained here showed a remarkable absorption enhancement with regards to the previously reported values of similar structures i.e. by Phin et al [13], or Choi et al [33]. On the other hand, the S-doped ZnO nanoflowers synthetized by Zhang's group [23], showed bigger band gap narrowing, than compared to the value of the $ZnO_{1-X}S_X$@CNTs obtained here. This may indicate that S in our case was not incorporated in high amount in the ZnO structure and/or we obtained just a thin layer of ZnS covering a ZnO core.

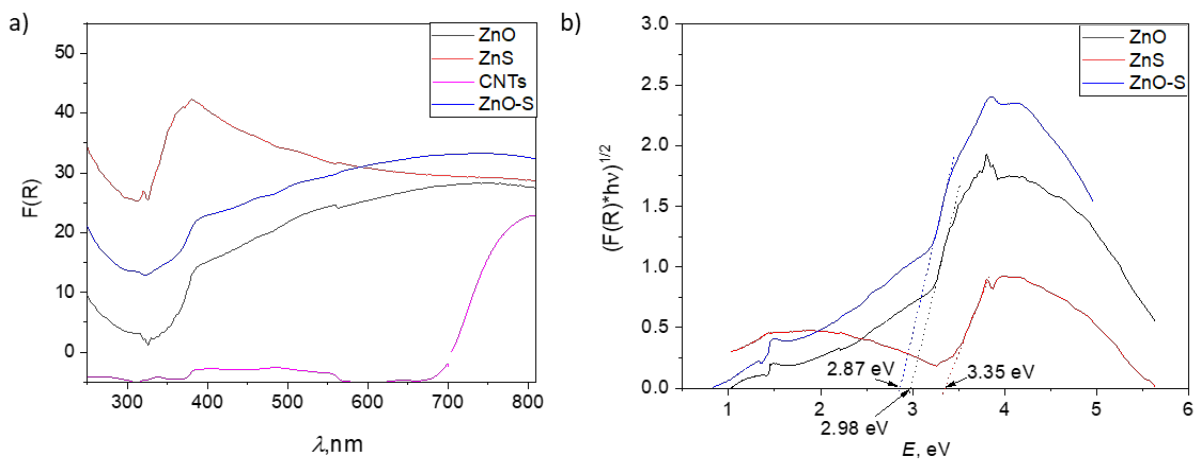

**Figure 8.** a) Kubelka-Munk function and b) Tauc plot obtained from DRS spectra of the bare CNTs film and CNTs heterostructures, where the spectrum of CNTs is given in violet, spectrum of ZnO/CNTs in black, ZnS/CNTs in red and S-ZnO/CNTs in blue

The photocatalytic activity of the three different heterostructured samples were compared and the results are summarized in Fig. 9. We can remark that for the monitored photolysis in absence of catalyst, the dye degraded at 27 % after 6 h, level of degradation that was close to the results using just CNTs film (not shown here). This value is attained with the use of any of the studied heterostructured catalyst 4-6 times faster, within 60-90 min. Furthermore, among the studied heterostructured catalysts, the $ZnO_{1-x}S_x$/CNTs showed the best efficiency by degrading the dye at 80 % in 6 h, whereas ZnO/CNTs reached 68 % and ZnS/CNTs 72 % degradation respectively. The photodegradation kinetics follows the pseudo first order equation, $\ln(c_0/c_i) = kt$, where k is the rate constant, $c_0$ and $c_i$ are the initial and final concentration of the MB dye at time 't', respectively and the rate constant plot is given in Fig. 9b. By applying linear fitting, the rate constants were calculated to be: $3.07 \times 10^{-3}$ min$^{-1}$ for ZnO/CNTs, $3.70 \times 10^{-3}$ min$^{-1}$ for ZnS/CNTs and $4.63 \times 10^{-3}$ min$^{-1}$ for $ZnO_xS_{1-x}$/CNTs catalysts.

We assume that S doping and/or creation ZnS outerlayer of ZnO give benefit to the photocatalytic performance of pure ZnO since under the light exposure, the photogenerated electrons pass from the conduction band of ZnS to the corresponding band of ZnO and hole transfer occurs from the valence band of ZnO to that of ZnS. The simultaneous transfer of electrons and holes in the ZnO–ZnS system should increase both the yield and the lifetime of charge carriers, in which the photogenerated electrons can be further captured by CNTs, whereas the holes attack the dye molecules. The photocatalysis efficiency is therefore significantly enhanced. This mechanism is proposed by Lin et al. [39]. Yu et al. also reported that the $O_2$ adsorbed on the surface of CNTs could accept photogenerated electron from the catalyst and in the presence of H$^+$ immediately forms a very reactive OH that eliminates organic molecules of a pollutant [40].

Herein we present the effective way to obtain a direct and stable contact between the deposit and CNTs by applying the non-thermal plasma treatment of the as-synthetized CNTs [41]. This was important to ensure efficient electron transfer from the photoactive material to CNTs [42]. It is to note that an optimization of the PED parameters was performed in order to assure a good dispersion of the deposit without agglomeration as this is prerequisite to ensure the high specific surface for photocatalytic reactions. Additionally, the electron-hole recombination which would otherwise happen at the grain boundaries of an agglomerated catalyst is decreased. The transfer of photoexcited electron from the catalyst to CNTs is expected to be faster than the recombination [43].

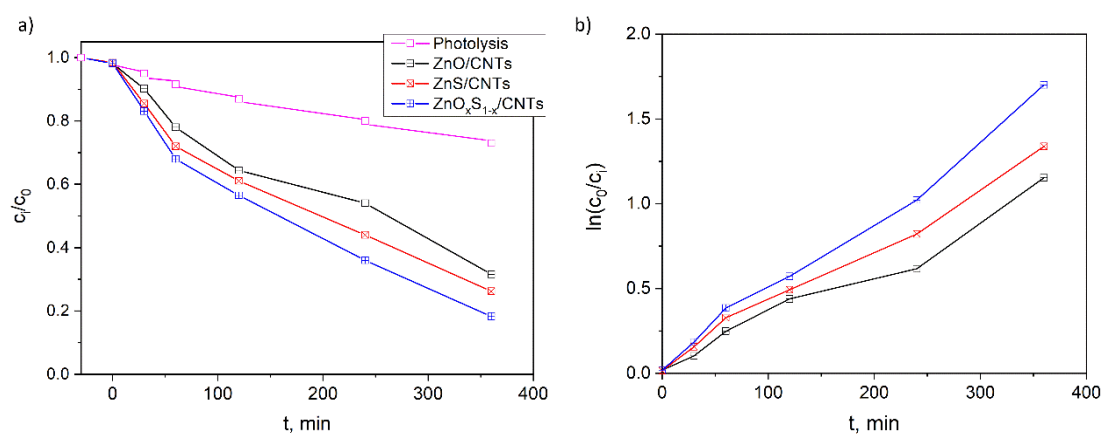

**Figure 9.** Photodegradation kinetics ($c_i/c_0$ vs. time) and b) Rate constant plot ($\ln(c_0/c_i) = kt$) of the different heterostructured samples.

## 4    Conclusion

Photocatalytic active ZnO1-xSx@CNTs heterostructures have been synthesized using a bottom-up process with vertically aligned carpets of surface modified CNTs as support for Zn nanoparticles electrodeposition and subsequent oxidation/sulfurization of the later. By combining controlled CNT's outer wall point defects creation and optimization of the electrodeposition parameters, we succeed to obtained well dispersed nanoparticles deposition all along CNTs' outer walls. The HRTEM study evidenced the presence of hexagonal ZnO along the CNTs, whereas the XRD analysis was consistent with a mixed deposit of pure Zn and a small amount of hexagonal ZnO. We can thus assume that ZnO is a cover shell layer of the electrodeposited Zn particles. Subsequent annealing in $H_2O$ vapors successfully yielded a pure ZnO hexagonal phase with no notable particles' morphology change. ZnS@CNTs heterostructure was obtained by the annealing of Zn@ZnO/CNTs starting material in S vapor. The composition of the deposit was investigated by XRD analysis and the HRTEM/EDS measurements that confirmed that hexagonal ZnS was obtained. With the aim to produce ZnO1-xSx@CNTs heterostructures, the effect of a two-steps annealing, first in $H_2O$ and subsequently in S vapors, was also investigated. The benefit of the low-temperature annealing in S vapor and obtaining a ZnO1-xSx@CNTs heterostructures was evidenced by DRS absorption analysis that highlighted the narrowing of the bang gap to

2.87 eV. It is also worth noticing that due to the CNTs, all the three types of heterostructures investigated exhibited an additional absorption red shift as compared to the values of absorption edges for pure ZnO and ZnS. This would mean that CNTs served as an optical sensitizer of the structures. The enhanced absorption properties are to be related to the measured improved photocatalytic behavior since the ZnO1-xSx@CNTs heteronanostructure showed the best efficiency by decolorizing the dye at 80 % in 6 h, whereas ZnO@CNTs achieved degradation at 68 % and respectively at 72 % for ZnS@CNTs.

Further investigations about the nonthermal plasma treatment influence on the surface defects creation on CNTs and subsequent Zn and similar transition metals and their oxides are undergoing.

## Acknowledgements

This work received financial support from Campus France within the MOPGA (Make Our Planet Great Again) program. The authors acknowledge financial support from the French state managed by the National Research Agency under the Investments for the Future program under the references ANR-10-EQPX-50 pole NanoMAX and from Chaire "André Citroen" at Ecole Polytechnique. All TEM analysis were performed at CIMEX (Centre Interdisciplinaire de Microscopie électronique de l'X) which is gratefully acknowledged. This work is part of the NanoMaDe-3E Initiative.

## References

[1]    S.K. Kansal, N. Kaur, S. Singh, Photocatalytic degradation of two commercial reactive dyes in

aqueous phase using nanophotocatalysts, Nanoscale Res. Lett. 4 (2009) 709–716

[2] F. Huang, H. Tang, Y. Wang, J. Hou, Z. Liu, R.C. Massé, J. Tian, G. Cao, Hierarchical ZnO microspheres photoelectrodes assembled with Zn chalcogenide passivation layer for high efficiency quantum dot sensitized solar cells, J. Power Sources. 401 (2018) 255–262.

[3] B. Simović, D. Poleti, A. Golubović, A. Matković, M. Šćepanović, B. Babić, G. Branković, Enhanced photocatalytic degradation of RO16 dye using Ag modified ZnO nanopowders prepared by the solvothermal method, Process. Appl. Ceram. 11 (2017) 27–38.

[4] G. Wang, B. Huang, Z. Li, Z. Lou, Z. Wang, Y. Dai, M.H. Whangbo, Synthesis and characterization of ZnS with controlled amount of S vacancies for photocatalytic $H_2$ production under visible light, Sci. Rep. 5 (2015) 1–7.

[5] L. Yin, D. Zhang, D. Wang, X. Kong, J. Huang, F. Wang, Y. Wu, Size dependent photocatalytic activity of ZnS nanostructures prepared by a facile precipitation method, Mater. Sci. Eng. B Solid-State Mater. Adv. Technol. 208 (2016) 15–21.

[6] M. Mao, L. Jiang, L. Wu, M. Zhang, T. Wang, The structure control of ZnS/graphene composites and their excellent properties for lithium-ion batteries, J. Mater. Chem. A. 3 (2015) 13384–13389.

[7] W. Yu, J. Zhang, T. Peng, New insight into the enhanced photocatalytic activity of N-, C- and S-doped ZnO photocatalysts, Appl. Catal. B Environ. 181 (2016) 220–227.

[8] Y. Tang, J. Tian, T. Malkoske, W. Le, B. Chen, Facile ultrasonic synthesis of novel zinc sulfide/carbon nanotube coaxial nanocables for enhanced photodegradation of methyl orange, J. Mater. Sci. 52 (2017) 1581–1589.

[9] P. Spathis, I. Pouliost, The corrosion and photocorrosion of zinc and zinc oxide coatings, Corros. Sci. 37 (1995) 673–680

[10] C. M. Taylor, A. Ramirez-Canon, J. Wenk, D. Mattia, Enhancing the photo-corrosion Resistance of ZnO Nanowire Photocatalysts, J. Hazard. Mater. 378 (2019) 120799

[11] T. Mahvelati-Shamsabadi, E.K. Goharshadi, Photostability and visible-light-driven photoactivity enhancement of hierarchical ZnS nanoparticles: The role of embedment of stable defect sites on the catalyst surface with the assistant of ultrasonic waves, Ultrason. Sonochem. 34 (2017) 78–89.

[12] G. Wan, G. Wang, Microwave-assisted synthesis and enhanced photocatalytic activity of carbon nanotubes supported ZnS nanoparticles, J. Optoelectron. Biomed. Mater. 7 (2015) 85–91.

[13] H.Y. Phin, Y.T. Ong, J.C. Sin, Effect of carbon nanotubes loading on the photocatalytic activity of zinc oxide/carbon nanotubes photocatalyst synthesized via a modified sol-gel method, J. Environ. Chem. Eng. (2019) 103222

[14] S.A. Feng, J.H. Zhao, Z.P. Zhu, The manufacture of carbon nanotubes decorated with ZnS to enhance the ZnS photocatalytic activity, Xinxing Tan Cailiao/ New Carbon Mater. 23 (2008) 228–234.

[15] R. Paul, P. Kumbhakar, A.K. Mitra, Synthesis and study of photoluminescence characteristics of carbon nanotube/ZnS hybrid nanostructures, J. Exp. Nanosci. 5 (2010) 363–373.

[16] Z. Fang, Y. Fan, Y. Liu, Photochemical synthesis and photocatalysis application of ZnS/amorphous carbon nanotubes composites, Front. Optoelectron. China. 4 (2011) 121–127.

[17] A.B. Patil, K.R. Patil, S.K. Pardeshi, Ecofriendly synthesis and solar photocatalytic activity of S-doped ZnO, J. Hazard. Mater. 183 (2010) 315–323.

[18] J. Xiao, W. Huang, Y. Hu, F. Zeng, Facile in situ synthesis of wurtzite ZnS/ZnO core/shell heterostructure with highly efficient visible-light photocatalytic activity and photostability, J. Phys. D: Appl. Phys. 51 (2018) 075501

[19] G. Song, W. Li, Fabrication of hollow ZnO particles and its photocatalytic property by modifying of nano ZnS, J. Nanosci. Nanotechnol. 13 (2013) 1364–1367.

[20] X. Huang, M.-G. Willinger, H. Fan, Z. Xie, L. Wang, A. Klein-Hoffmann, F. Girgsdies, C.-S. Lee, X.-M. Meng, Single crystalline wurtzite ZnO/zinc blende ZnS coaxial heterojunctions and hollow zinc blende ZnS nanotubes: synthesis, structural characterization and optical properties, Nanoscale. 6 (2014) 8787–8795.

[21] S. Jeong, M. Choe, J.W. Kang, M.W. Kim, W.G. Jung, Y.C. Leem, J. Chun, B.J. Kim, S.J. Park, High-performance photoconductivity and electrical transport of ZnO/ZnS core/shell nanowires for multifunctional nanodevice applications, ACS Appl. Mater. Interfaces. 6 (2014) 6170–6176.

[22] Y. Hu, H. Qian, Y. Liu, G. Du, F. Zhang, L. Wang, X. Hu, A microwave-assisted rapid route to synthesize ZnO/ZnS core-shell nanostructures via controllable surface sulfidation of ZnO nanorods, CrystEngComm. 13 (2011) 3438–3443.

[23] L. Zhang, X. Zhu, Z. Wang, S. Yun, T. Guo, J. Zhang, T. Hu, J. Jiang, J. Chen, Synthesis of ZnO doped high valence S element and study of photogenerated charges properties, RSC Adv. 9 (2019) 4422–4427.

[24] K.-H. Kim, A. Gohier, J.E. Bourée, M. Châtelet,

C.-S. Cojocaru, The role of catalytic nanoparticle pretreatment on the growth of vertically aligned carbon nanotubes by hot-filament chemical vapor deposition, Thin Solid Films. 575 (2015) 84–91.

[25] A. Castan, S. Forel, L. Catala, I. Florea, F. Fossard, F. Bouanis, A. Andrieux-Ledier, S. Mazerat, T. Mallah, V. Huc, A. Loiseau, C.S. Cojocaru, New method for the growth of single-walled carbon nanotubes from bimetallic nanoalloy catalysts based on Prussian blue analog precursors, Carbon N. Y. 123 (2017) 583–592.

[26] Z.B. He, J.L. Maurice, C.S. Lee, A. Gohier, D. Pribat, P. Legagneux, C.S. Cojocaru, Etchant-induced shaping of nanoparticle catalysts during chemical vapour growth of carbon nanofibres, Carbon N. Y. 49 (2011) 435–444.

[27] B. Marquardt, L. Eude, M. Gowtham, G. Cho, H.J. Jeong, M. Châtelet, C.S. Cojocaru, B.S. Kim, D. Pribat, Density control of electrodeposited Ni nanoparticles/nanowires inside porous anodic alumina templates by an exponential anodization voltage decrease, Nanotechnology. 19 (2008) 405607

[28] Langford, J.I., The Use of the Voigt Function in Determining Microstructural Properties from Diffraction Data by means of Pattern Decomposition, in: Accuracy Powder Diffr. II, Gaithersburg, National Institute of Standards and Technology Special Publication, Gaithersburg, 1992: p. 846.

[29] F. and T.I. Izumi, Implementation of the Williamson–Hall and Halder–Wagner Methods into RIETAN-FP, Adv Inst Sci Technol. 3 (2014) 33–38.

[30] H. B. Ahn, J. Y. Lee, High-resolution transmission electron microscopy study of twinned ZnS nanoparticles, Mater. Lett. 106 (2013) 308–312.

[31] C. C. Cheng, W. C. Weng, H. I. Lin, J. L. Chiu, H. Y. Jhao, Y. T.A. Liao, C. T.R. Yu, H. Chen, Fabrication and characterization of distinctive ZnO/ZnS core-shell structures on silicon substrates: Via a hydrothermal method, RSC Adv. 8 (2018) 26341–26348.

[32] A. Bjelajac, V. Djokic, R. Petrovic, G. Socol, I.N. Mihailescu, I. Florea, O. Ersen, D. Janackovic, Visible light-harvesting of TiO₂ nanotubes array by pulsed laser deposited CdS, Appl. Surf. Sci. 309 (2014) 225–230.

[33] M. S. Choi, T. Park, W.J. Kim, J. Hur, High-performance ultraviolet photodetector based on a zinc oxide nanoparticle@single-walled carbon

nanotube heterojunction hybrid film, Nanomaterials. 10 (2020).

[34] S. P. Lonkar, V. V. Pillai, S.M. Alhassan, Facile and scalable production of heterostructured ZnS-ZnO/Graphene nano-photocatalysts for environmental remediation, Sci. Rep. 8 (2018) 1–14.

[35] R. R. Thankalekshmi, A.C. Rastogi, Structure and optical band gap of $ZnO_{1-x}S_x$ thin films synthesized by chemical spray pyrolysis for application in solar cells, J. Appl. Phys. 112 (2012) 1–10.

[36] L. Zhao, L. Gao, Coating multi-walled carbon nanotubes with zinc sulfide, J. Mater. Chem. (2004) 1001–1004.

[37] M. J. Casciato, G. Levitin, D.W. Hess, M.A. Grover, Synthesis of optically active ZnS-carbon nanotube nanocomposites in supercritical carbon dioxide via a single source diethyldithiocarbamate precursor, Ind. Eng. Chem. Res. 51 (2012) 11710–11716.

[38] A. Del Gobboa, Silvano Mottrama, S. Ould-Chikhb, J. Chaopaknama, P. Pattanasattayavonga, V. D'Elia, A. Valerio, Physico-chemical investigation of ZnS thin-film deposited from ligand- free nanocrystals synthesized by non-hydrolytic thio-sol-gel, Nanotechnology. 29 (2018) 385603.

[39] D. Lin, H. Wu, R. Zhang, W. Zhang, W. Panw, Facile synthesis of heterostructured ZnO-ZnS nanocables and enhanced photocatalytic activity, J. Am. Ceram. Soc. 93 (2010) 3384–3389.

[40] Y. Yu, J.C. Yu, C.Y. Chan, Y.K. Che, J.C. Zhao, L. Ding, W.K. Ge, P.K. Wong, Enhancement of adsorption and photocatalytic activity of TiO₂ by using carbon nanotubes for the treatment of azo dye, Appl. Catal. B Environ. 61 (2005) 1–11.

[41] K. H. Kim, D. Brunel, A. Gohier, L. Sacco, M. Châtelet, C.S. Cojocaru, Cup-stacked carbon nanotube schottky diodes for photovoltaics and photodetectors, Adv. Mater. 26 (2014) 4363–4369.

[42] J. Chang, J.-H. Lee, C.K. Najeeb, J.-H. Kim, Surface-Enhanced Raman Scattering of Carbon Nanotubes by Decoration of ZnS Nanoparticles, J. Nanosci. Nanotechnol. 11 (2011) 6253–6257.

[43] H. K. Sharma, S.K. Sharma, K. Vemula, A.R. Koirala, H.M. Yadav, B.P. Singh, CNT facilitated interfacial charge transfer of TiO₂ nanocomposite for controlling the electron-hole recombination, Solid State Sci. 112 (2021) 106492.

# SUPPLEMENTARY INFORMATION

## Photocatalytic active $ZnO_{1-x}S_x$/CNTs heterostructures

Andjelika Bjelajac[a, b*], Ileana Florea[a, c], Mihai Zamfir[a, d], Sandrine Tusseau Nenez[e] and Costel Sorin Cojocaru[a]

[a]LPICM, CNRS, Ecole polytechnique, IP Paris, 91228 Palaiseau Cedex, France.

[b]Luxembourg Institute of Science and Technology (LIST), Maison des Matériaux, 28, avenue des Hauts-Fourneaux, L-4365 Esch-sur-Alzette, Luxembourg.

[c]CRHEA, CNRS, UMR7073, Rue Bernard Grégory, 06905 Sophia-Antipolis Cedex, France.

[d]National Institute for Laser, Plasma & Radiation Physics (INFLPR), Atomistilor Street, No. 409, Magurele, Ilfov RO-077125, Romania

[e]LPMC, CNRS, Ecole polytechnique, IP Paris, 91228 Palaiseau Cedex, France

[*]Corresponding author, andelika.bjelajac@list.lu

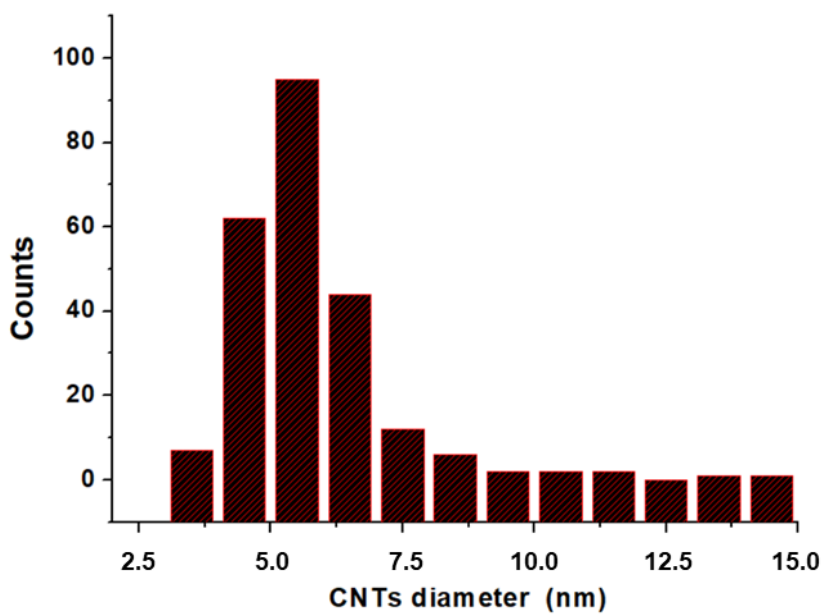

Fig. S1. Diameter size distribution histogram of the pristine CNTs

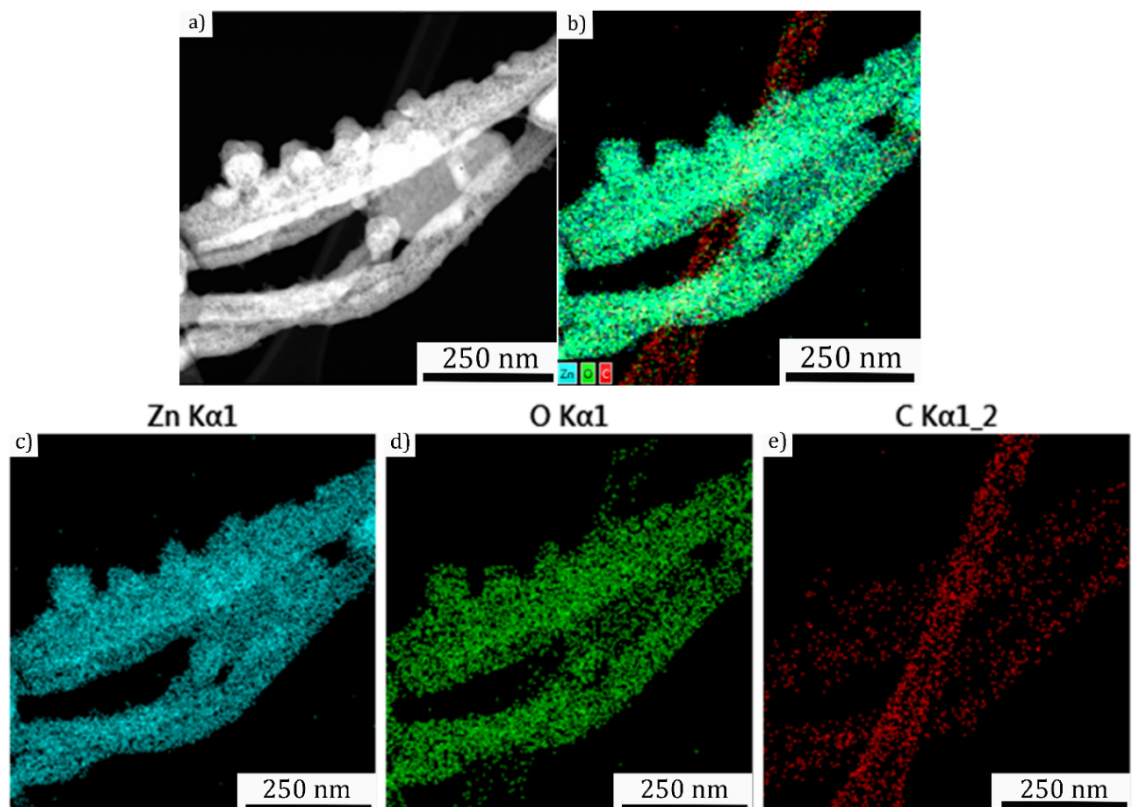

Fig. S2. STEM-HAADF EDS chemical analysis of an area specific to the Zn@ZnO/CNTs structure: a) STEM-HAADF image and b) its corresponding chemical mapping with the Zn component in blue (c), the O in green (d) and the C component in red (e)

Table S1. Chemical composition of the analyzed fragment presented in Fig. S1, obtained by EDS analysis (Cu originates from the TEM grid)

| Element line | C Kα | O K | Si K | Cu K | Zn K | Total |
|---|---|---|---|---|---|---|
| At. % | 19.5 | 46.7 | 1.1 | 2.6 | 30.1 | 100.0 |

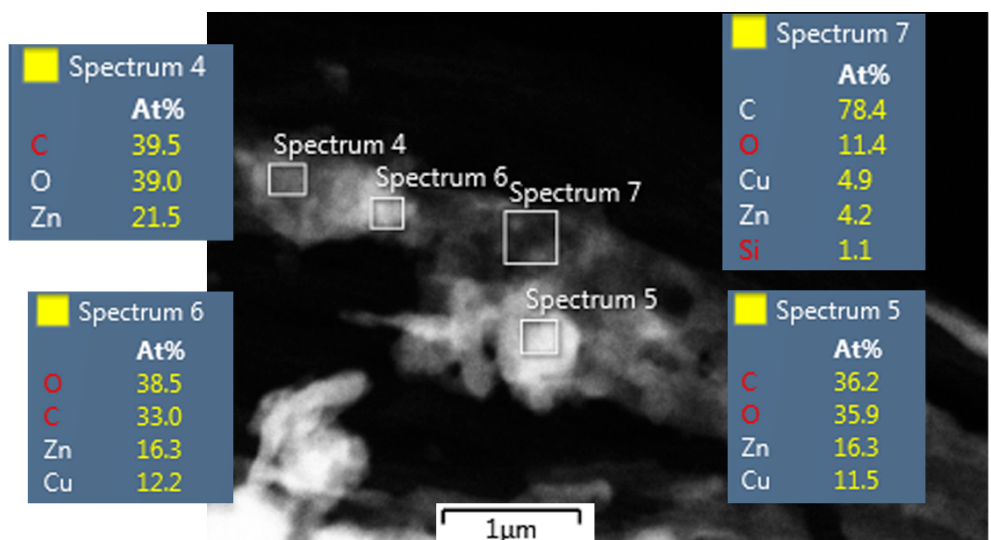

Fig. S3 EDS point analysis of the Zn as deposited on CNTs

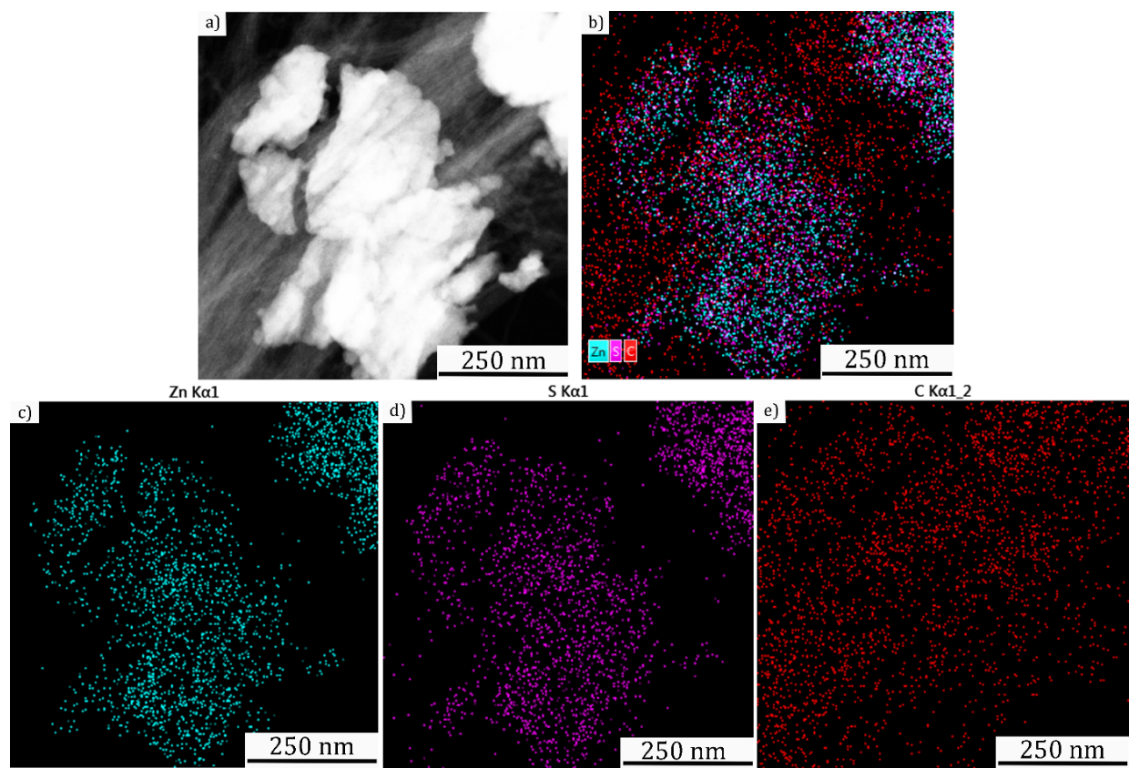

Fig. S4. STEM-HAADF EDS chemical analysis of an area specific to the ZnS/CNTs structure: a) STEM-HAADF image and b) its corresponding chemical mapping with the Zn component in blue c), the S in violet d), and the C component in red e)

Table S2. Chemical composition of the analyzed fragment presented in Fig. 6, obtained by EDS analysis

| Element line | C Kα | O K | Si K | Cu K | Zn K | S K | Total |
|---|---|---|---|---|---|---|---|
| At. % | 92.5 | 0.9 | 0.6 | 0.6 | 2.4 | 3.1 | 100.0 |

Table S3. Chemical composition of the analyzed fragment presented in Fig. 7, obtained by EDS analysis

| Element line | C Kα | O K | Si K | Cu K | Zn K | S K | Total |
|---|---|---|---|---|---|---|---|
| At. % | 62.7 | 10.4 | 2.0 | 3.8 | 15.2 | 5.9 | 100.0 |